\pdfoutput=1
\documentclass[sigconf]{acmart}
\usepackage{booktabs} 
\usepackage{xcolor}

\usepackage{times}  
\usepackage{hyperref}
\usepackage{enumerate}

\usepackage{blindtext}
\usepackage[T1]{fontenc}
\usepackage[utf8]{inputenc}
\usepackage{newtxmath,bm,courier}
\usepackage{xspace,enumitem,fancyhdr,fancyref,wrapfig}
\usepackage{amsmath,amsfonts,bm}
\usepackage{tabularx,tablefootnote,wrapfig,hyphenat}
\usepackage[labelformat=simple,skip=0pt]{subcaption}
\usepackage{xfrac,sparklines}
\usepackage{algorithm}
\usepackage{algpseudocode}

\usepackage{tabularx}
\usepackage{wrapfig}
\usepackage{lipsum}
\usepackage{mathtools}
\usepackage{amsmath}
\usepackage{upgreek}
\usepackage{float}

\usepackage[toc]{appendix}

\newcommand{\shortname}{mmWall}
\newcommand{\systemname}{\shortname{}}
\newcommand{\shortnames}{mmWall's}

\setitemize{itemsep=3pt,topsep=4pt,parsep=2pt,partopsep=0pt,leftmargin=1.5em}
\setenumerate{itemsep=3pt,topsep=4pt,parsep=1pt,partopsep=0pt,leftmargin=1.5em}

\copyrightyear{2021}
\acmYear{2021}
\setcopyright{acmcopyright}\acmConference[HotMobile '21]{The 22nd International
Workshop on Mobile Computing Systems and Applications}{February 24--26,
2021}{Virtual, United Kingdom}
\acmBooktitle{The 22nd International Workshop on Mobile Computing Systems and
Applications (HotMobile '21), February 24--26, 2021, Virtual, United Kingdom}
\acmPrice{15.00}
\acmDOI{10.1145/3446382.3448665}
\acmISBN{978-1-4503-8323-3/21/02}

\begin{document}
\title{mmWall: A Reconfigurable Metamaterial Surface for mmWave Networks}

\author{Kun Woo Cho\texorpdfstring{$^1$}{}, Mohammad H. Mazaheri\texorpdfstring{$^2$}{}, Jeremy Gummeson\texorpdfstring{$^3$}{}, Omid Abari\texorpdfstring{$^4$}{}, Kyle Jamieson\texorpdfstring{$^1$}{}}
\affiliation{Princeton Univ.\texorpdfstring{$^1$}{}, Univ. of Waterloo\texorpdfstring{$^2$}{}, Univ. of Massachusetts Amherst\texorpdfstring{$^3$}{}, UCLA\texorpdfstring{$^4$}{}}



\begin{abstract}

To support faster and more efficient networks, 
mobile operators and service providers are bringing
5G millimeter wave (mmWave) networks indoors. However, due to their high 
directionality, mmWave links are extremely vulnerable to
blockage by walls and human mobility. 
To address these challenges, 
we exploit advances in artificially-engineered metamaterials, introducing
a wall-mounted smart metasurface, called \textit{\shortname{}}, that enables a fast mmWave beam relay 
through the wall and 
redirects the beam power
to another direction
when a human body blocks a line-of-sight path.
Moreover, our \shortname{} supports multiple users and fast beam alignment by generating multi-armed beams.
We sketch the design of a real-time system by considering (1) how to design 
a programmable, metamaterials-based surface
that refracts the incoming signal to one or more arbitrary directions, and
(2) how to split an incoming mmWave beam into multiple outgoing beams 
and arbitrarily control the beam energy between these beams.
Preliminary results show the \shortname{} metasurface steers the outgoing 
beam in a
full $360$-degrees,  with an $89.8\%$ single-beam efficiency 
and $74.5\%$ double-beam efficiency.
\end{abstract}

\ccsdesc[500]{Hardware~Analysis and design of emerging devices and systems}
\ccsdesc[500]{Hardware~Wireless devices}

\keywords{Metamaterials, Metasurface, Reconfigurable Intelligent Surfaces, mmWave Networks, Split Ring Resonators}
\hypersetup{pdfauthor={Name}}

\maketitle


\begin{figure*}[t]
\begin{subfigure}[a.]{0.24\linewidth}
\includegraphics[width=0.95\linewidth]{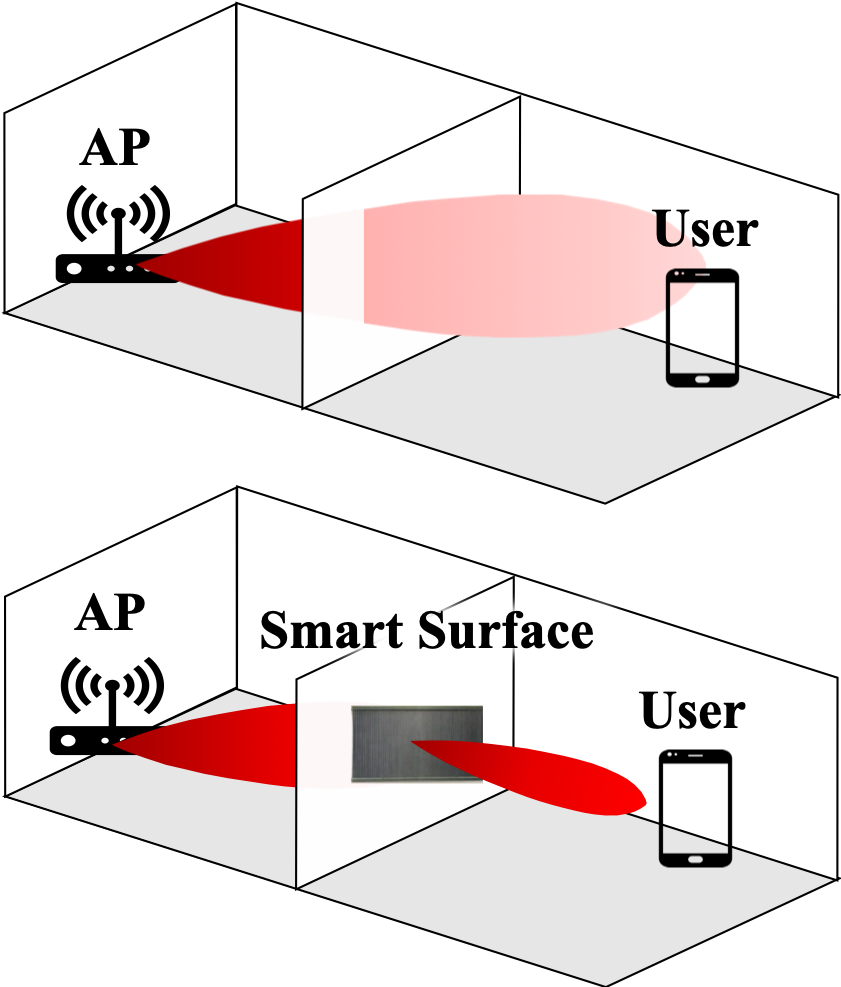}
\caption{Beam relay and focusing.}
\label{f:scenario1}
\end{subfigure}
\begin{subfigure}[b.]{0.24\linewidth}
\includegraphics[width=0.95\linewidth]{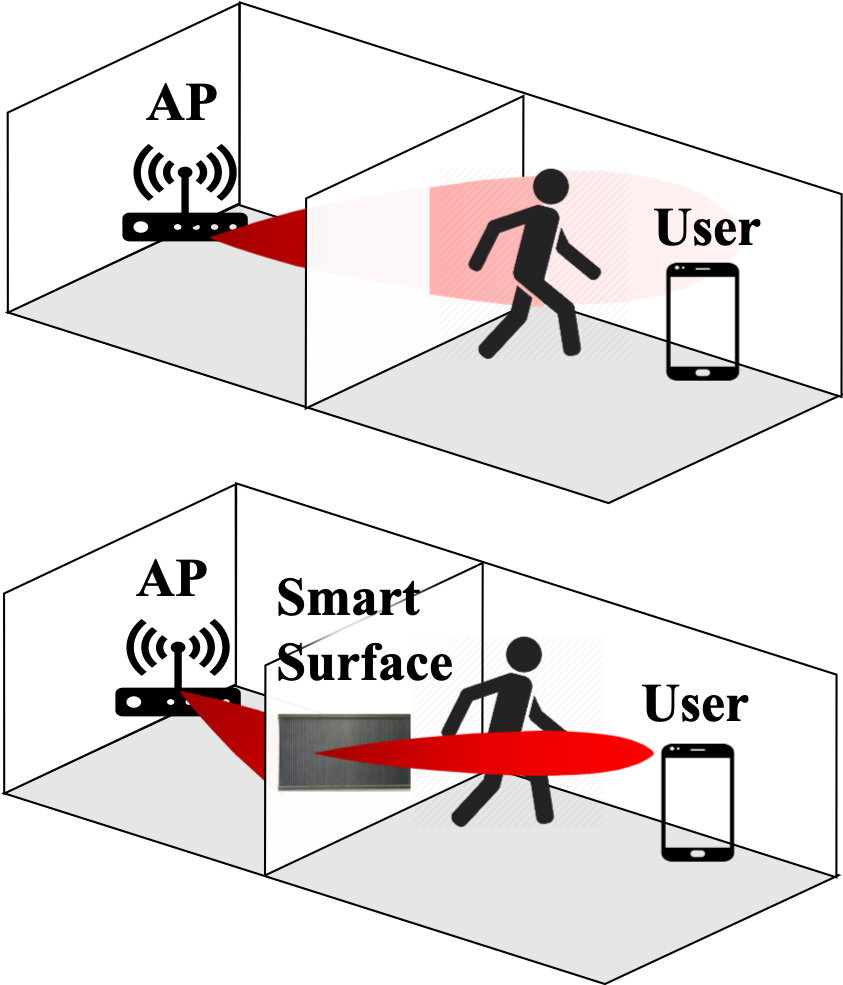}
\caption{Path diversity.}
\label{f:scenario2}
\end{subfigure}
\begin{subfigure}[b.]{0.24\linewidth}
\includegraphics[width=0.95\linewidth]{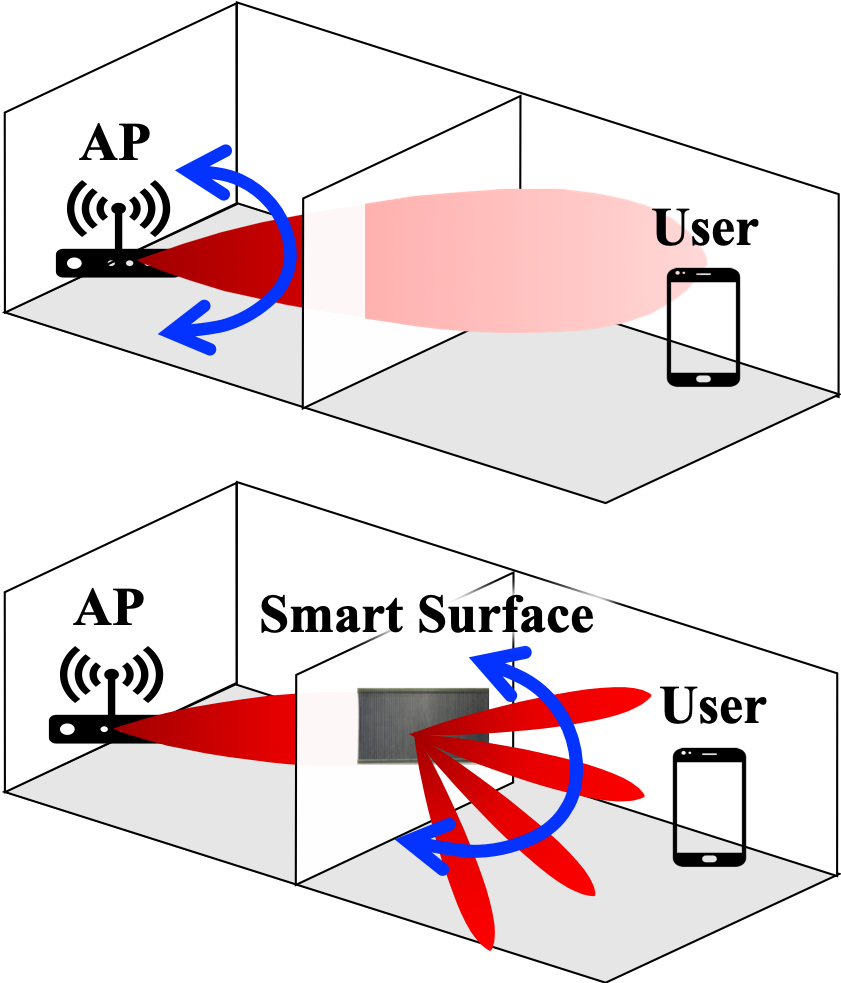}
\caption{Beam searching.}
\label{f:scenario3}
\end{subfigure}
\begin{subfigure}[b.]{0.24\linewidth}
\includegraphics[width=0.95\linewidth]{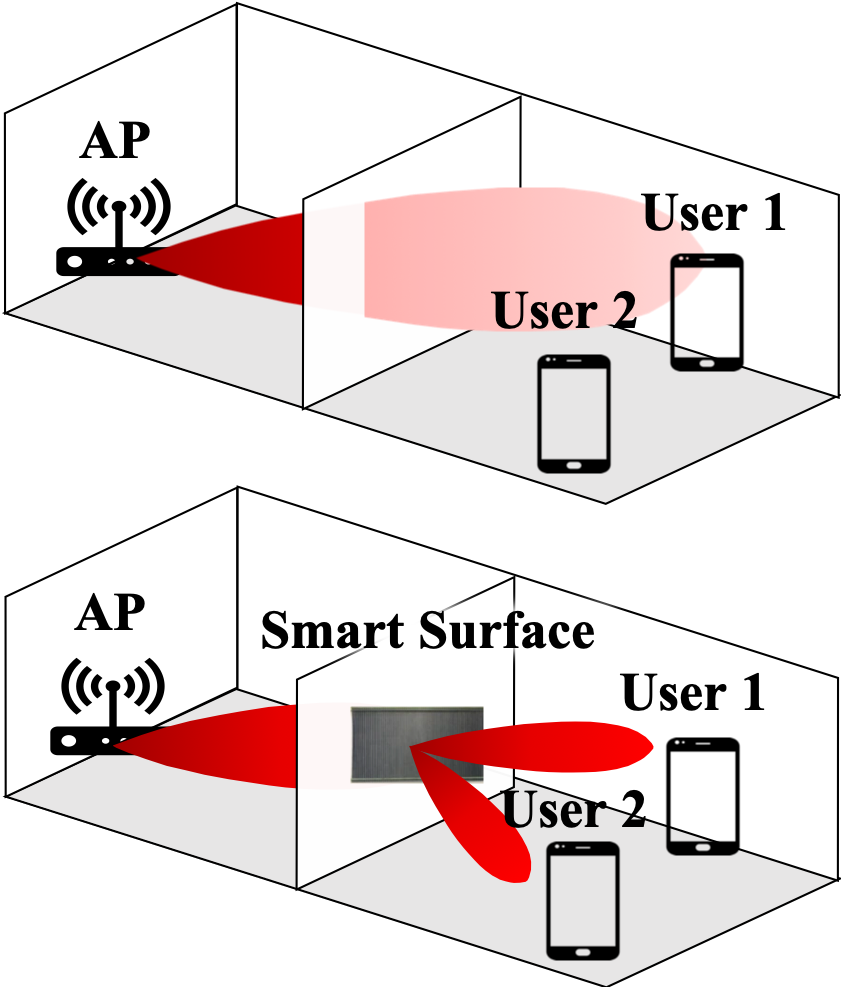}
\caption{Multicast.}
\label{f:scenario4}
\end{subfigure}
\caption{Illustration of \shortname{} use cases
(\emph{Upper:} without \shortname{},  \emph{Lower:} with \shortname{}).
(a) without \shortname{}, the beam power attenuates as it traverses the wall and its width widens. With \shortname{}, we relay and focus the beam towards the user; (b) when a human body blocks the signal, \shortname{} redirects the beam power to another direction to provide an alternative signal path; (c) \shortname{} generates multiple beams to probe the best direction between the user and AP in a timely manner; (d) \shortname{} splits the incoming beam to simultaneously support multiple users.}
\label{f:example}
\vspace{-6pt}
\end{figure*}

\section{Introduction}
\label{s:intro}

The use of millimeter-wave
(mmWave) spectrum has emerged in the 5G era as a key next generation wireless 
network technology to fulfill users' demands for high spectral efficiency and low
latency wireless networks. 
Higher carrier frequencies offer greater network capacity: for instance, 
the maximum carrier frequency of the 4G-LTE band at 2.4~GHz provides an 
available spectrum bandwidth of only 100~MHz, while mmwave 
(above 24~GHz) can easily hold spectral bandwidths five to ten times greater,
enabling multi-Gbit/second data rates. 
This way, mmWave 
enables a plethora of mobile, wireless applications, such as
VR/AR for multiplayer games,
camera-tracking in smart stores, and robotic automation in smart warehouses,
that are currently infeasible due to their requirements of high 
bandwidth.

However, 
mmWave technology faces a big challenge due to its
weak ability to diffract around obstacles with a size significantly larger than the wavelength. 
Since mmWave has an extremely short wavelength, it experiences huge losses when traversing walls, and thus requires a line-of-sight (LoS) path
between the transmitter and receiver. 
Moreover, to compensate for the propagation loss, 
mmWave uses highly directional antennas to focus the signal power in a narrow beam. 
Since the mmWave beam is very narrow, communication glitches occur whenever humans walk across such ``pencil-beam'', resulting in a significant SNR drop of $20$ dB~\cite{abari2017enabling}.

One na\"ive solution is to deploy multiple mmWave AP in every room to guarantee the LoS communication. However, this not only increases the cost of mmWave implementation, but also incurs huge complexity in coordinating a massive number of mmWave APs, especially in the presence of human mobility.
Instead of simply increasing the endpoints of the wireless links and 
leaving the wireless channel itself unchanged, we ask a following question: can we build a smarter radio environment, one that electronically reconfigures itself to relay the mmWave beam?

To answer this question, we propose a \textbf{Reconfigurable \emph{Metasurface for mmWave Network} (\shortname{})}, a tunable smart surface made of metamaterial. 
Unlike a conventional wireless relay system, our reconfigurable metasurface does not have transmitting and receiving antennas, nor an amplifier. Once the incoming beam hits the metasurface, it naturally refracts the beam into a desired direction, regardless of whether the transmitter and receiver are located in the same room ("mirror" mode) or in the different room ("lens" mode). Also, it can split the incoming signal into multiple beams and concurrently steer the multi-armed beams.

As illustrated in Fig.~\ref{f:example}, we can summarize the key advantages of \shortname{} as follow:
(1) \shortname{} relays and re-focuses the beam at the wall. This way, the beam does not attenuate as it traverses the wall;
(2) \shortname{} provides an alternative path when the direct path is blocked by a human body.
(3) Due to a narrow beam, the current solutions for aligning the beams scan the entire space to find the best direction, which incurs a large time delay. By generating the multi-armed beams with different spatial directions and sweeping them concurrently, \shortname{} enables the fast beam search at the wall; 
(4) \shortname{} enables multi-cast by creating and directing one beam per each user;
(5) 
Each element of metasurface, known as a meta-atom, is 
at least five times smaller than the conventional antenna. 
\shortname{} thus 
has a larger number of elements
than
the phased array antenna 
and yields a significant gain enhancement;
(6) 
In general, multiple RF chains are used to generate multiple beams, 
  resulting in a huge power consumption and complex hardware design.
  In contrast, \shortname{} doesn't have a single RF chain and therefore intrinsically consumes low-power. 


\noindent
\textbf{Roadmap.} 
Section~\ref{s:theoretical} theoretically analyzes Huygens' metasurface at a finer level. 
Section~\ref{s:design} introduces a novel approach to scale Huygens' metasurface to higher mmWave frequency
and elaborate how this metasurface relays one or more beams with a full-angle coverage of $360$ degrees.
In Section~\ref{s:eval} we then conduct a preliminary study to prove the efficiency of beam relay 
and steering performance.
Section~\ref{s:related} presents the related works, Section~\ref{s:applications} addresses the mobile applications of \shortname{}, and Section~\ref{s:discussion} concludes the paper with discussion.


\section{Primer: Huygens Metasurfaces}
\label{s:theoretical}
\begin{figure}[t]
\centering
\includegraphics[width=\linewidth]{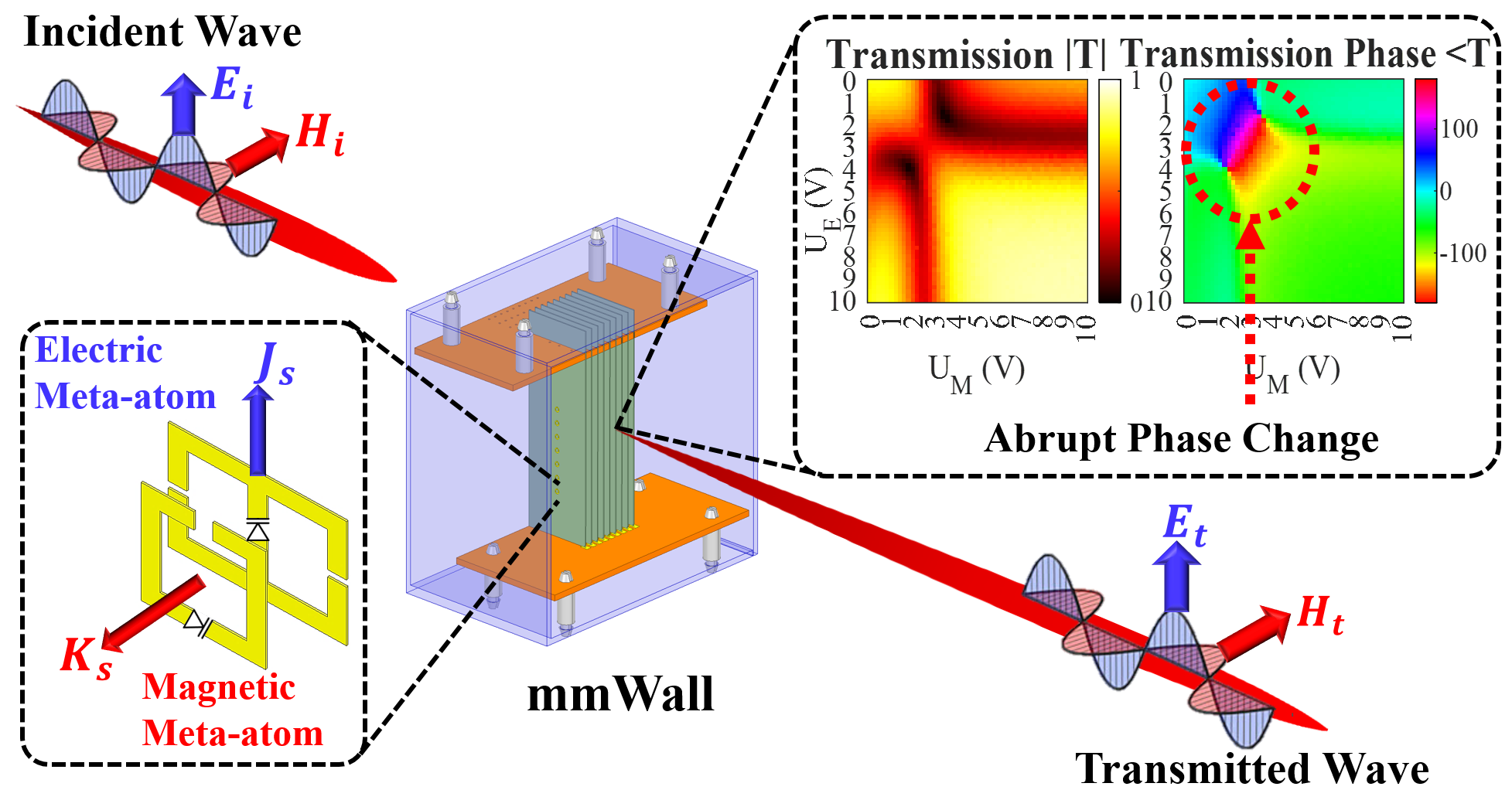}
\caption{Diagram of a Huygens' metasurface. An incident wave ($\vec{E_{i}},\vec{H_{i}}$) is converted to a transmitted wave ($\vec{E_{t}},\vec{H_{t}}$) through a field discontinuity, sustained by electric and magnetic currents densities ($\vec{J_{s}},\vec{K_{s}}$), induced by the electric and magnetic meta-atoms. \textit{Bottom-left:} the electric and magnetic meta-atom pair; \textit{upper-right:} the Huygen's pattern with varying voltage ($U_{E}$,$U_{M}$). We apply $U_{E}$ and $U_{M}$ to varactor on the electric and magnetic meta-atom, respectively.}
\label{f:huygen_surface}
\end{figure}

Huygens' metasurfaces (HMSs)~\cite{ding2020metasurface, liu2018huygens, chen2017reconfigurable, zhang2018space} comprise a layer of co-located orthogonal electric and magnetic meta-atom,  
facing each other across dielectric substrate (See metallic rings in Fig.~\ref{f:huygen_surface}). 
This meta-atom pair introduces a discontinuity in the electromagnetic fields and hence providing the means for manipulation of all attributes of the incident field, including its magnitude and phase. 
Specifically, as the incident wave ($\vec{E_{i}},\vec{H_{i}}$) passes through the magnetic meta-atom, the magnetic field $\vec{H_i}$ of the incident wave induces the rotating current within the metallic loop of the magnetic meta-atom that in turn produces its own magnetic field $\vec{K_s}$, which enhance or oppose the incident field. 
Similarly, the electric meta-atom
is excited by the electric field $\vec{E_i}$ of the incident wave, resulting in the oscillating
current loops that creates its own electric response $\vec{J_s}$.
When these electric and magnetic response of the Huygens' meta-atom pair interact with the fields of the incident wave, it creates an abrupt phase shift. 
Hence, by controlling the electric and magnetic responses, Huygens' metasurface steers the incoming wave to a desired direction.
For readers interested in the detailed description refer to~\cite{chen2018huygens}.

\noindent
\textbf{Theoretical Analysis. }
To induce the magnetic or electric response, the magnetic and electric meta-atom each acts as a resonant $LC$ circuit, 
a circuit consisting of an inductor $L$ and a capacitor $C$. 
Figure~\ref{f:designs} shows the design parameters of the Huygens' meta-atoms. 
The gap of a metallic loop induces a capacitance 
while the metallic loop itself generates an inductance.
Specifically, we can simplify the equation of a gap capacitance as:
    $C = \epsilon_{0} wt/g$
where $\epsilon_{0}$ is free-space permittivity with the the length of the gap $g$, width of the metallic ring $w$, and metal thickness $t$. 
The simplified equation of inductance is written as:
   $L = \mu_{0} l_{1}l_{2}/t$
where $\mu_{0}$ is free-space permeability, and $l_{1}$ and $l_{2}$ are the dimension of the meta-atom. 
Since HMS is engineered to oscillate at a resonant frequency $f_{0}=1/(2\pi \sqrt{LC})$, 
we must modify $LC$ value to design HMS at a desired frequency.
Consider a $LC$ circuit with one gap capacitor and one inductor (\textit{i.e.} magnetic meta-atom).
We can simplify the expression for the $LC$ circuit as follow:
\begin{equation}
\vspace{-6pt}
\begin{array}{l}
     LC = (\mu_{0} A/t) (\epsilon_{0} wt/g) =(1/c_{light}^2)*(Aw/g)\\
\end{array}
\label{eq:lc}
\end{equation}
where $c_{light}$ is the speed of light, and $A$ is the area of meta-atom, which is equivalent to $l_{1}l_{2}$ in Fig.~\ref{f:old_huygen}. 

\noindent
\textbf{Active HMS.}
To render HMS reconfigurable, tunable electrical components, 
such as varactors and p-i-n diodes, are loaded to the meta-atom 
so that it can tune a metasurface element through voltage bias lines.
For \shortname{}, varactor is used as a voltage-controlled capacitor.
As seen in Fig.~\ref{f:analysis}, this varactor capacitor $C_{var}$ 
forms a series circuit with the gap capacitor $C_{gap}$.
Assuming the equivalent circuit diagram of the magnetic meta-atom,
the total capacitance can be written as:
$C_{total} = \frac{1}{(1/C_{gap})+(1/C_{var})}$. 
Hence, by applying voltages across varactor, 
we can slightly shift the resonant frequency from the designed frequency $f_{target}=1/(2\pi \sqrt{LC_{gap}})$ to $f_{new}=1/(2\pi \sqrt{LC_{total}})$. Therefore we provide different transmission phases
at the targeting frequency $f_{target}=1/(2\pi \sqrt{LC_{gap}})$. 
The upper-right graphs in Fig.~\ref{f:huygen_surface} shows the amplitude of the transmission coefficient $|T|$ and the phase of the transmission coefficient $<T$ with varying voltage $U_{E}$ and $U_{M}$ for the electric and magnetic meta-atoms, respectively. This pattern is called Huygens' pattern, and it has a full transmission-phase coverage of $360$ degrees with a high amplitude.

\section{Design}
\label{s:design}

\shortname{} is a programmable metasurface that operates at 
mmWave frequency, fully controls the direction of the mm\-Wave 
beam, and splits the relayed mmWave beam into multiple directions. 
In this section, we describe our solutions to the challenges 
in designing the \shortname{}. 

\subsection{\shortname{} meta-atom design}
\begin{figure}[t]

\begin{subfigure}[a.]{.32\linewidth}
\includegraphics[width=.84\linewidth]{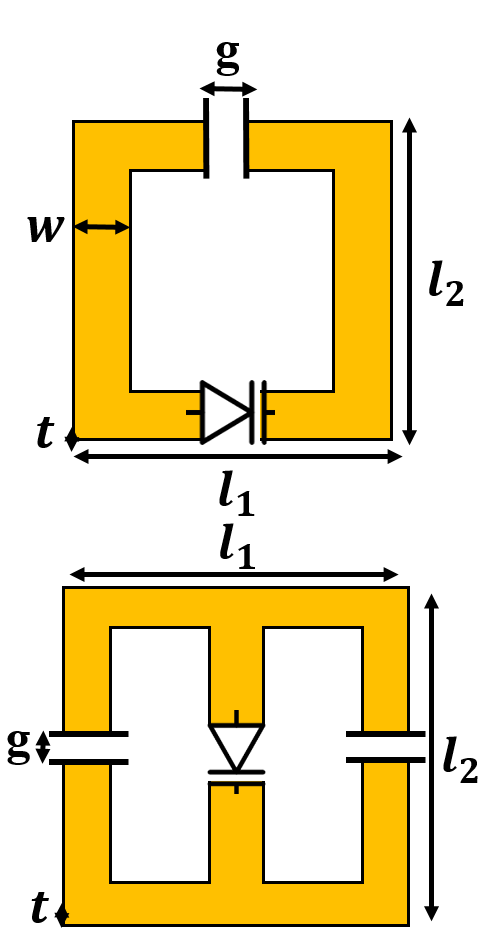}
\caption{Huygens design.}
\label{f:old_huygen}
\end{subfigure}
\begin{subfigure}[a.]{.32\linewidth}
\includegraphics[width=.84\linewidth]{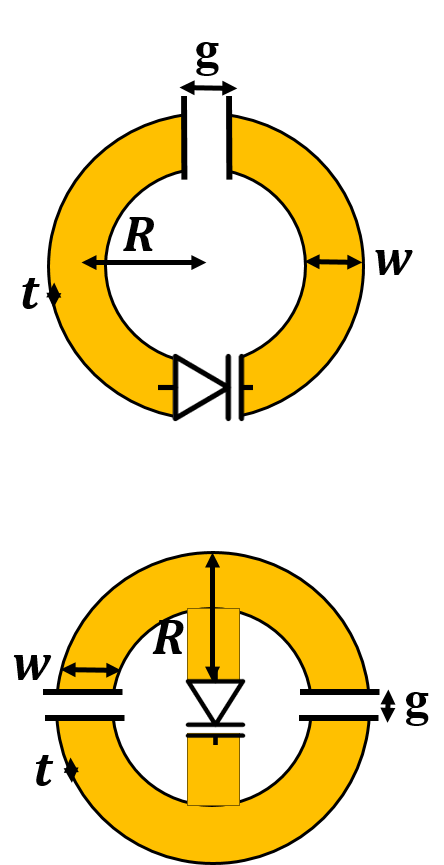}
\caption{\systemname{} design.}
\label{f:new_huygen}
\end{subfigure}
\begin{subfigure}[b.]{.332\linewidth}
\includegraphics[width=.855\linewidth]{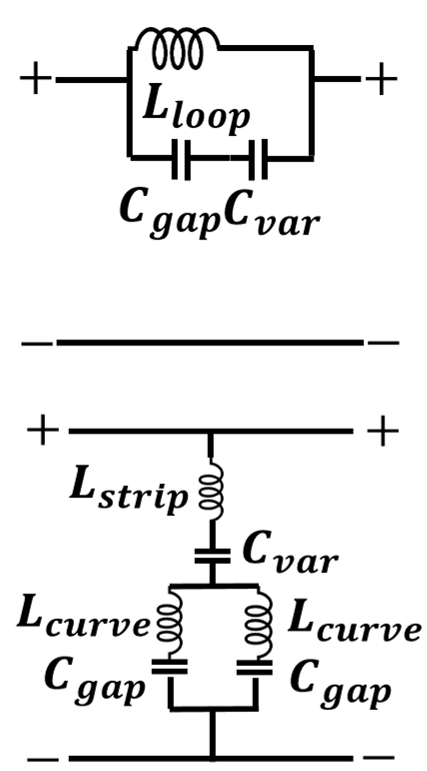}
\caption{Circuit diagram.}
\label{f:analysis}
\end{subfigure}
\caption{The original and \systemname{} meta-atom design and the equivalent circuit diagram of both designs: for each of 
(a), (b) and (c), top is the magnetic side, and the bottom is the electric side. $L_{curve}$ indicates the inductance of the copper strips on left or right. $L_{strip}$ stands for the inductance of a single copper strip on the middle of the electric meta-atom.}
\label{f:designs}
\vspace{-5pt}
\end{figure}

To design \shortnames{} meta-atoms to be effective at mmWave frequencies,
we need to redesign the meta-atom such that its targeting resonant frequency $f_{target}$ matches the mmWave frequency at $24$ GHz. 
Given the Eq.~\ref{eq:lc},
we can
increase the targeting frequency to $24$ GHz by 
decreasing the area $A=l_{1}l_{2}$, ring width $w$, and/or increasing the gap size $g$. 
Accordingly, the simplest, conventional solution is to directly scale down the size of the meta-atom design, 
such that each $l_{1}$ and $l_{2}$ equals $\lambda/10$, which is a standard meta-atom size. 
At mmWave, however, a meta-atom with $\lambda/10$ size is too small such that once we load a smallest-available varactor, 
its packaging completely distorts the tailored electromagnetic surface properties.
Hence, we instead reduce the area $A$ by changing the rectangular meta-atom into a circular design with the radius $R=l_{1}/2=l_{2}/2$.
To load varactor, we 
then fix $l_{1}$ and $l_{2}$ to twice the size of varactor (\textit{i.e.} breaking the $\lambda/10$ convention). 
New meta-atom design for \shortname{} is illustrated in Fig.~\ref{f:new_huygen}. 
With this new meta-atom design, we reduce the area by a factor of $\pi/4$, which in turn lowers the inductance $L$ and increases the resonant frequency, given the resonant frequency equation.
Then, based on theoretical analysis we derived,
we fine-tune other geometric parameters until the targeting frequency $f_{target}=1/(2\pi \sqrt{LC_{gap}})$ reaches $24$ GHz. 

\subsection{\systemname{} relay beam-steering}
The conventional phased array antenna calculates the total field pattern
by multiplying the element factor, a pattern produced by a single element, to
the array factor, a pattern produced by an array of elements.
Consider array of $n$ identical antennas with $d$ spacing and amplitude $a$. 
The array factor is:
\begin{equation}\vspace{-4pt}
\begin{array}{l}
AF {=} a {+} ae^{jkd(cos\theta)} {+} \dots {+} ae^{jk(N-1)d(cos\theta)}
{=}a{\sum\limits_{n=0}^{N-1}} e^{jknd(cos\theta)}
\end{array}
\label{eq:array_factor}
\end{equation}
where $k=2\pi/\lambda$ with $\lambda$ as the wavelength of the operating frequency, and $\theta$ is the steering angle. 
As seen in Eq~\ref{eq:array_factor}, 
the phase shift of each element is different. 
More specifically, the phase of $n^{th}$ element is larger than the phase of element $n-1$ by
$kd cos\theta$, since the path length to $n^{th}$ element is 
$dcos\theta$ longer than $(n-1)^{th}$ element. 
Consequently, to steer the beam towards a particular direction, 
the phased array antenna must apply different phases for each array element, 
and the larger the phase difference is, the greater the phased array antenna steers.

Similarly, \shortname{} applies a different phase shift at each meta-atom to steer the beam. 
Specifically, we leverage a full $360$ degree transmission-phase coverage of HMS to provide different phase shifts.
Let's assume that we have an array of meta-atom pairs as shown in Fig.~\ref{f:beam_steer}. 
For the $1^{st}$ meta-atom pair, 
we apply $0V$ for the magnetic meta-atom and 
$0V$ for the electric meta-atom, such that ($U_{E},U_{M}$) = ($0V,0V$). 
When we directly map this voltage pair to the Hugyens' pattern in Fig.~\ref{f:beam_steer}, 
the $1^{st}$ meta-atom pair provides a transmission-phase of $\phi$ with a high transmission-amplitude $a_0$. 
For the $2^{nd}$ meta-atom pair, 
$1V$ is applied to both meta-atoms, resulting in a different phase shift, $2\phi$, with a high transmission amplitude $a_1$ and so on.
At last, we can formulate the array factor as follow:
\begin{equation}
\begin{array}{l}
AF = a + a_{1}e^{j\phi} + \dots +a_{n}e^{j(N-1)\phi}
\approx a\sum\limits_{n=0}^{N-1} e^{jn\phi}
\end{array}
\label{eq:array_factor_hms}
\end{equation}
This equation corresponds to Eq.~\ref{eq:array_factor}, and 
therefore we can steer the beam using HMS by varying the voltages. 
We must note that 
while the existing relay systems 
requires two phase antenna arrays, one to receive the incoming signal and another to transmit a new signal with a time delay, 
our \shortname{} only need a single array of the meta-atoms as it directly shift the phase of the existing signal. 

\subsection{\shortname{} beam splitting and searching}

\begin{figure}[t]
\begin{subfigure}[a.]{.545\linewidth}
\includegraphics[width=\linewidth]{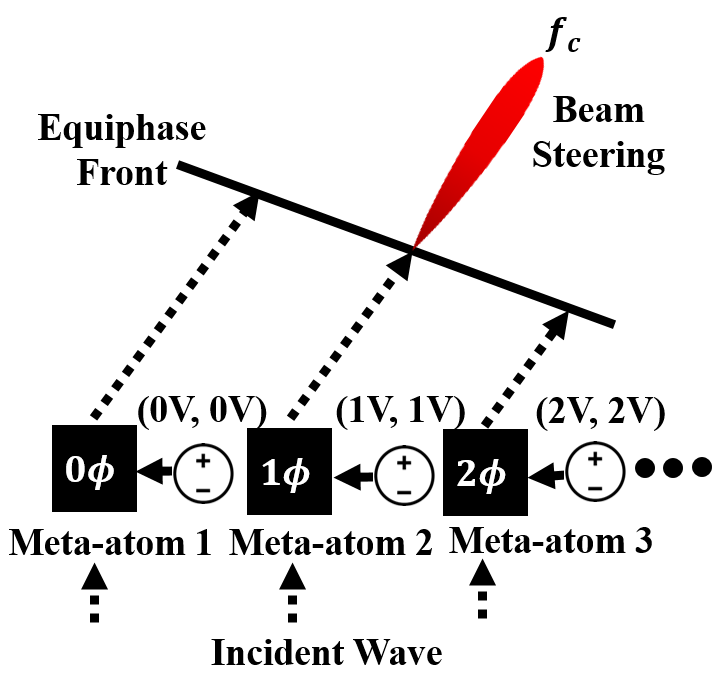}
\end{subfigure}
\begin{subfigure}[b.]{.445\linewidth}
\includegraphics[width=\linewidth]{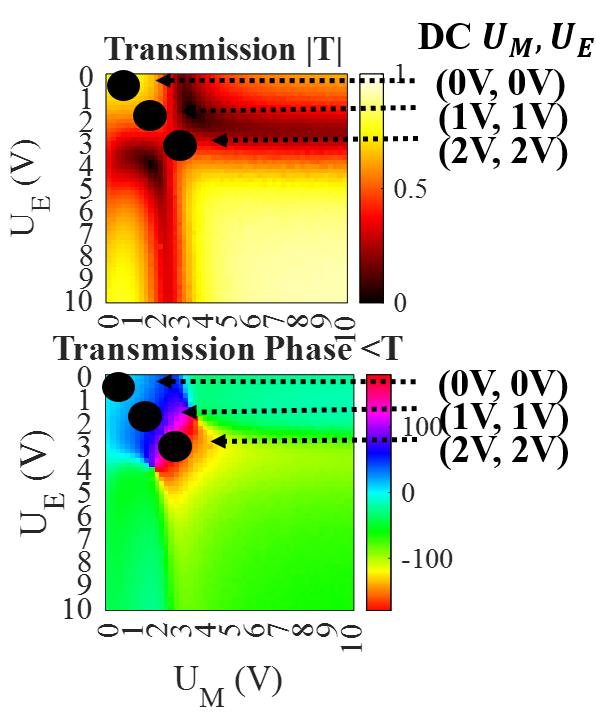}
\end{subfigure}
\vspace{-10pt}
\caption{
Schematic illustration 
of a beam control using \shortname{}.
The voltages ($U_{E},U_{M}$)
for each meta-atom pair 
on left is mapped and visualized with Huygens' pattern on right.}
\label{f:beam_steer}
\end{figure}



To transform an incoming beam into multi-armed beams, 
\shortname{} adopts an additional degree-of-freedom, \textit{time}, in the control line. 
In other words, we convolve a high frequency mmWave signal with a low frequency signal and 
thus achieve a desired Fourier series. 
More specifically, we add time-modulation in the voltage signal 
to achieve a time-varying transmission-amplitude |T| and phase <T. 
When this time-varying transmission signal is periodic, 
its Fourier transformation
becomes harmonics 
and creates multiple beams with different frequencies, also known as \textit{sidebands}. 
Hence, by applying a proper time-varying voltage signal, 
\shortname{} can generate the time-varying signal response, 
in which its Fourier transformation creates a desired number of beams at desired frequencies.
Let us define each of time-varying voltage signal $\tilde{U_{E}}$ and 
$\tilde{U_{M}}$ for the electric and magnetic meta-atom as $U_{amp}f(t)+U_{offset}$ where
$U_{amp}$ is the voltage amplitude, $U_{off}$ is the voltage offset, and
$F(t)=\sum_{n}^{N}a^{(n)}cos[n(\Omega t-\varphi]+b^{(n)}sin[n(\Omega t-\varphi]$ is a normalized Fourier series 
with a modulation frequency $\Omega$, time $t$, and phase $\varphi$.
Our goal is to find the solution $\Uptheta^*$ to the following optimization problem:
\begin{equation}
 \Uptheta^* = \underset{\Uptheta}{\arg\max}~OBJ(\mathscr{F}(T(\Uptheta)))
\end{equation}
where $\Uptheta^*$ is an optimal set of the voltage waveform coefficients ($U_{amp}$,$U_{off}$,$\left \{ a \right \}$,$\left \{ b \right \}$,$\varphi$). $T(\Uptheta)$ is a 
mapping function from the voltage waveform
to the Huygens' pattern in Fig.~\ref{f:beam_steer}, and
$\mathscr{F}(T(\Uptheta))$ is a Fourier transformation of the time-varying transmission signal. 
\begin{figure}[t]
\begin{subfigure}[a.]{.553\linewidth}
\includegraphics[width=\linewidth]{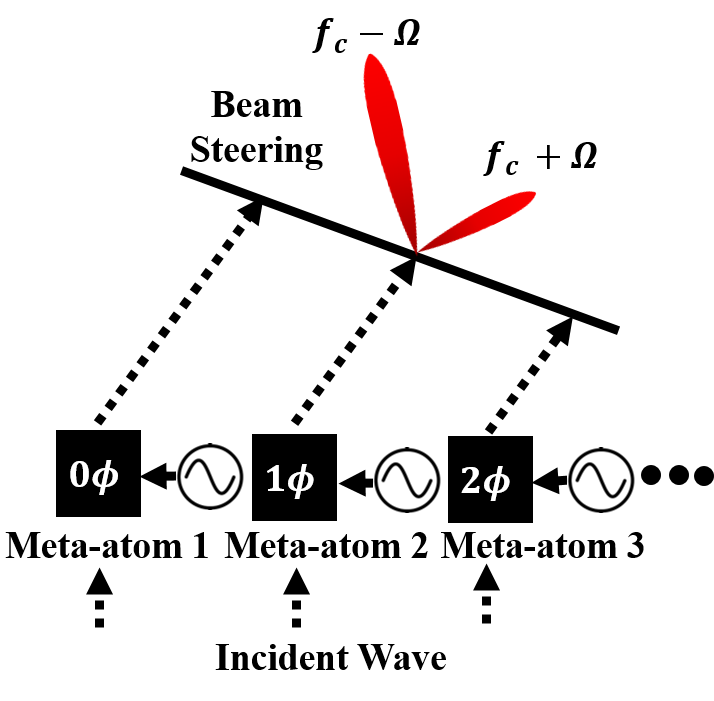}
\end{subfigure}
\begin{subfigure}[b.]{.4 \linewidth}
\includegraphics[width=\linewidth]{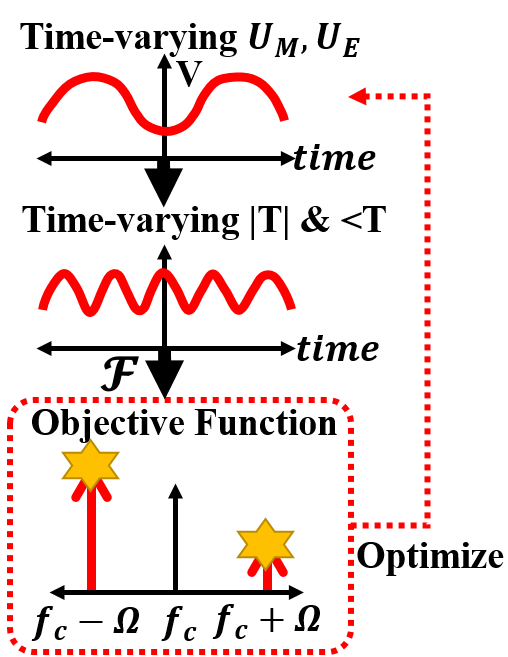}
\end{subfigure}
\vspace{-10pt}
\caption{Schematic illustration of multi-armed beam generation at $f_c-\Omega$ and $f_c+\Omega$ and multi-armed beam-steering. 
$f_c$ is the carrier frequency, and $\Omega$ is the voltage modulation frequency.}
\label{f:multibeam_steer}
\end{figure}
Finally, $OBJ(\mathscr{F}(T(\Uptheta)))$ is the objective function characterizing the scattered power of the desired beams at a desired frequency bin.

\noindent
\textbf{Example.}
In Fig.~\ref{f:beam_steer}, we apply a constant DC voltage 
to each meta-atom,
resulting in a constant transmission-amplitude |T| and phase <T that creates a single beam at a carrier frequency $f_{c}$. 
Consider the multi-armed beams scenario in Fig.~\ref{f:multibeam_steer}.
Its goal is to split the beam into two, one at $f_{c}+\Omega$ and 
another at $f_{c}-\Omega$ 
where $\Omega$ is the voltage modulation frequency. 
Let's assume that we want to concentrate more energy towards the beam at $f_{c}-\Omega$ than the beam at $f_{c}+\Omega$.
We can redefine our goal to search for a proper $\Uptheta$
, such that $\mathscr{F}(T(\Uptheta))$ has a large peak at $f_{c}-\Omega$ and relatively low peak at $f_{c}+\Omega$. 
Accordingly, in Fig.~\ref{f:multibeam_steer}, our objective function is to maximize the sum of the weighted power at $f_{c}-\Omega$ and at $f_{c}+\Omega$, as denoted with the stars. 
After optimizing with the genetics algorithm, 
\shortname{} concurrently steer the multi-armed beams 
by applying different phase shifts to the optimized voltage waveform of each meta-atom pair, 
as seen in Fig.~\ref{f:multibeam_steer}. 

\noindent
\textbf{Mirror Mode. }
\shortname{} can also reflect the signal back as a mirror. 
To convert the transmissive mode to the reflective mode, we can simply add $180$ degrees phase shift to either $\tilde{U_{E}}$ or $\tilde{U_{M}}$. When the phase of $\tilde{U_{E}}$ and $\tilde{U_{M}}$ is identical, the \shortname{} acts as a "lens" whereas with $180$ degree phase difference between $\tilde{U_{E}}$ and $\tilde{U_{M}}$, our \shortname{} acts as a "mirror".

\begin{figure*}[t]
\vspace{-5pt}
\begin{subfigure}[a.]{0.33\linewidth}
\includegraphics[width=1\linewidth]{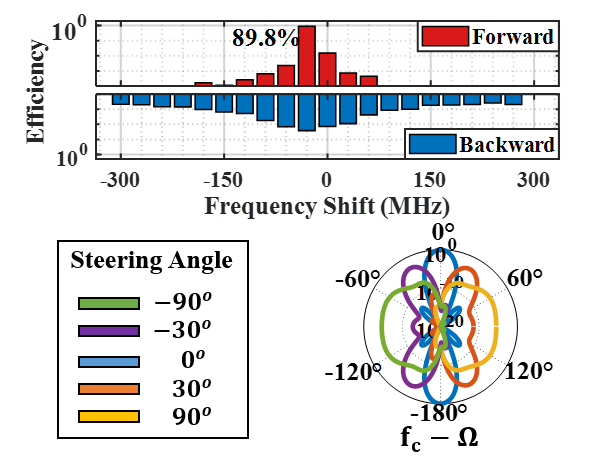}
\caption{Single transmissive beam}
\label{f:single}
\end{subfigure}
\begin{subfigure}[b.]{0.33\linewidth}
\includegraphics[width=1\linewidth]{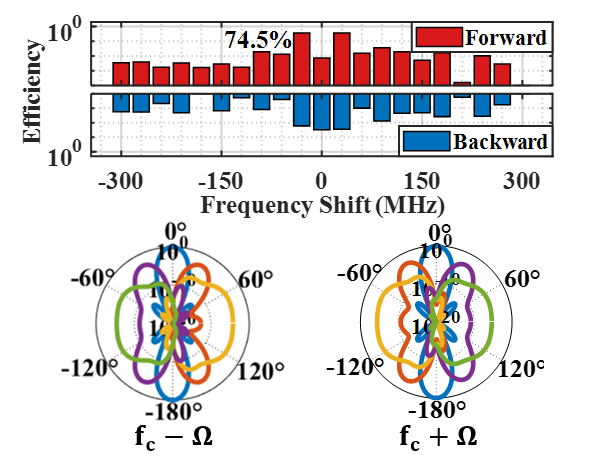}
\caption{Double transmissive beam}
\label{f:double_forward}
\end{subfigure}
\begin{subfigure}[a.]{0.33\linewidth}
\includegraphics[width=1\linewidth]{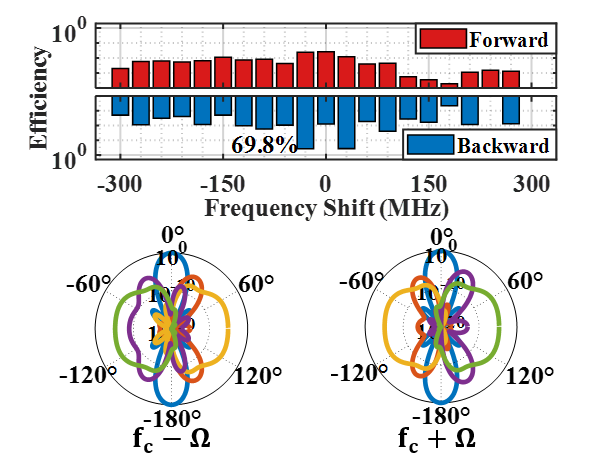}
\caption{Double reflective beam}
\label{f:douible_backward}
\end{subfigure}
\caption{\emph{Upper:} Beam efficiency versus frequency shift in log scale.  \emph{Lower:} beam-steering accuracy.
}
\label{f:eval}
\vspace{-7pt}
\end{figure*}

\begin{figure}[b]
\vspace{-10pt}
\includegraphics[width=\linewidth]{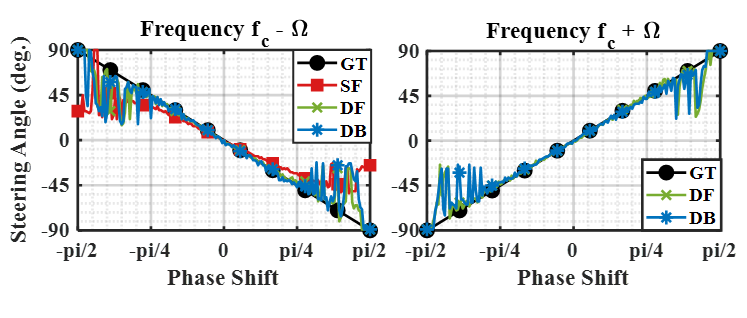}
\caption{Steering angle accuracy. GT, SF, DF, and DB stand for groundtruth, single forward, double forward, and double backward, respectively.}
\label{f:angle_accuracy}
\end{figure}

\section{Preliminary Results}
\label{s:eval}
To evaluate \shortname{}'s ability to relay and steer the mmWave beams, we ran HFSS simulation on our \shortname{} design with $20$ meta-atoms. 
We also modeled each electric component based on its Simulation Program with Integrated
Circuit Emphasis (SPICE) model.
We use mmWave signals at $24$~GHz and 
set the voltage modulation frequency $\Omega$ to $30$ MHz. 
We evaluate three scenarios: 
single transmissive beam, 
two transmissive beams, 
and two reflective beams. 
For a single beam, we transform the incident wave at $f_c$ to the signal at $f_{c}-\Omega$ 
whereas for the multi-armed beam scenarios, we translate the incoming beam into two beams, 
one at $f_{c}-\Omega$ and another at $f_{c}+\Omega$.

\noindent
\textbf{Beam Efficiency.}
Figure~\ref{f:eval} demonstrates the efficiency of \shortname{}. 
For a single transmissive beam (Fig.~\ref{f:single}), 
we observe a single peak at the $-30$ MHz frequency shift of the forward spectrum. 
Here, the frequency shift of $0$ MHz indicates carrier frequency at $24$ GHz, 
and $-30$ MHz frequency shift denotes $24$ GHz - $30$ MHz, which is equivalent to $f_{c}-\Omega$. 
This peak has $89.8\%$ efficiency, 
indicating that nearly $90\%$ of the incident signal 
is relayed with only $-0.46$ dB loss.
For the double transmissive beams and double reflective beams, 
the sum of beam efficiency at $f_{c}-\Omega$ and $f_{c}+\Omega$ is 
$74.5\%$ (approx. $1.28$ dB loss) and $69.8\%$ (approx. $1.56$ dB loss), accordingly. 
Furthermore, as we steer the beam away from $0$ degree 
where $0$ degree means no phase shift applied across the meta-atoms, 
the peak power of the beam weakens, slightly.
For all three scenarios, there is about $-2$ dB loss 
as we steer the beam by $-30$ or $30$ degree. 
At $-90$ or $90$ degree of the beam-steering angle, 
there is approximately $-3$ dB signal loss.  
We must further note that for the double-beam scenarios, 
\shortname{} correctly steers the beam at $f_{c}-\Omega$ and another beam at $f_{c}+\Omega$ in different directions, simultaneously.

\noindent
\textbf{Beam Steering Accuracy.}
Figure~\ref{f:angle_accuracy} evaluates the beam steering performance of three scenarios
as we vary the relative phase value $\phi$ across different meta-atoms. 
We specifically measure the angle of the peak of the beam.
The graph on left shows the steering performance of the beam at $f_{c}-\Omega$ for all three scenarios, 
and the graph on right illustrates steering accuracy of the beam at $f_{c}+\Omega$ for the double beam scenarios. 
The ground-truth values are colored in black with the circle markers. 
For every scenario, the beam steering angle is highly accurate except for the angle below $-\pi/4$ phase shift and above $\pi/4$ phase shift. 
This is because as it steers to a larger angle, the beam may not look symmetric, and thus peak may not be in its center of the beam. 

\section{Related Work}
\label{s:related}

\textbf{Smart Surfaces.}
LAIA~\cite{li2019towards} 
helps endpoints by
collecting radio energy from
one side of the wall, phase-shifting it, and transmitting it to the
other side such that the signals arriving in different paths combine constructively. 
RFocus~\cite{arun2020rfocus} is 
a software-controlled surface, 
made up of thousands of switching elements to focus reflections from a
transmitter to a receiver.
However, both LAIA and RFocus operate
at $2.4$ GHz that has much smaller path loss. 
Hence, their goal is intrinsically different from ours.
First, they don't address the issue of beam directionality since the beam at $2.4$ GHz is not narrow. 
Instead, they simply change the phase of the destructively interfering signals to improve overall signal strength. 
Second, since the effect of blockage is negligible, they don't consider the human blockage. 
MoVR~\cite{abari2017enabling} is a programmable mmWave reflector consists of two directional phased-array antennas.
When the LoS signal between the AP and VR headset is blocked by human mobility,
the AP steers its beam towards the MoVR, which in turn reflects it towards the headset. 
However, 
MoVR only works in a reflective "mirror" mode. In contrast, our \shortname{} 
can both relay the beam to another room 
and/or reflect it back to the same room.
Furthermore, 
unlike \shortname{}, MoVR cannot generate multiple beams.

\noindent
\textbf{Reconfigurable Metasurfaces.}
Artificially-engineered surfaces, called “meta-surfaces”, have been studied extensively by applied physicists to explore unique electromagnetic properties that do not exist in naturally occurring material. 
Recent work has proposed metasurfaces that can alter existing signals
in the environment itself, such as creating materials with
a negative refraction index~\cite{smith2004metamaterials}, engineering complex
beam patterns, and manipulating the signal polarization~\cite{chen2020pushing, wu2019tunable}.  
Especially, Huygens' metasurface (HMS) has gained an attention as a new paradigm for beam refraction, beamforming, and perfect reflection due to its key features: a full transmission-phase coverage of $360$ degrees and near-unity transmission. 
Unlike HMS that manipulates incoming waves in a passive manner, ample works have proposed the active-controlled Huygens' metasurfaces~\cite{chen2017reconfigurable, zhang2018space, liu2018huygens} by 
loading a tunable electric component like varactors.
While 
these designs have shown a great promise in controlled experiments that quantify performance at a low frequency (approx. $4$ to $6$ GHz), we cannot directly scale them to higher frequencies due to the conflict between the required meta-atom size 
and the size of varactor. 
Furthermore, they have not been integrated into an end-to-end system
for wireless communication.
In contrast, we 
present a system that directly improve wireless communication. 
\section{Mobile Applications}
\label{s:applications}
We describe the potential mobile, wireless applications that \shortname{} can support in real-time and significantly improve.

\noindent
\textbf{Virtual and/or Augmented Reality.}
We have witnessed major advances in virtual reality (VR), augmented reality (AR) and 
mixed reality (MR) in the gaming and entertainment industry and educational institutions. 
For instance, VR and AR can not only provide immersive gaming experiences, 
but also help in guiding the curated objects in a museum with digital versions of artists
and enable life-like training simulations to prepare public safety professionals.  
However, 
such applications are currently limited in terms of mobility as they require 
a physical connection via HDMI cable to exchange
multiple Gbps of data between a data source (PC or game console) and the headset.
For this reason, multiple companies have advocated
the use of mmWave links for such applications, 
which, in turn, experiences a significant difficulty 
with the presence of obstacles and reflections. 
Several mmWave relay systems~\cite{abari2017enabling,tan2018enabling} have attempted to solve this link-blockage problem 
using reconfigurable mmWave reflectors that provide an alternative path 
when the existing links are blocked. 
However, these systems only allow mobility within a single room as they can only reflect the signal back. 
Our \shortname{}, on the other hand, adaptively establishes a robust mmWave connection 
through both a reflective 
and transmissive path across the wall, 
thus enhancing the ability to move around through live events, with better sense of "presence".

\noindent
\textbf{Serverless Computing.}
Serverless computing is a cloud computing system in which the end users
run applications without a traditional server operating system. 
Instead, the service operators provide and manage machine resources on demand. 
The most prominent platforms include Amazon Web Service (AWS), Google Cloud, Microsoft Azure, and Cloudflare.
In particular, Google Cloud recently launched Game Servers, a managed service that provides gamers 
a cloud backend for running their games, including multi-player games.
Such services open up a plethora of computing opportunities for mobile devices,
which are often limited in computational resources.
However, the strict latency constraints exist when the operators support real-time services like Game Servers,
and these constraints exacerbate when the end users are mobile.
With \shortname{}, we can significantly reduce this latency by continuously supporting multiple Gbps
regardless of whether the users walk across the offices or not.
Furthermore, \shortname{} operates on both the downlink and the uplink, 
which are necessary for such services that require bi-directional communication.  
Thereby, \shortname{} can improve the Quality of Service (QoS) for the mobile devices using serverless computing. 

\noindent
\textbf{Robotic Automation.}
Robotic automation requires high speed connectivity 
to stream video to the backend servers in order to accomplish the complex collaborative tasks.
In fact, mmWave networks can play a significant role in providing high speed connectivity. 
However, when it comes to the smart robotic wearhouses and retailors,
there are an enormous amount of end-nodes to support.
Deploying multiple mmWave AP in every corner and space 
may allow multiple Gbps connection through LoS paths
but with a massive number of mobile robots, it adds additional complexity to the handover process. 
The use of \shortname{} not only mitigates complexity in coordinating the communication between the APs and the robots, 
but also
leaves the end-node unchanged. 
Only the AP and \shortname{} need to be configured so that \shortname{} delegates the task of the AP.

\section{Discussion}
\label{s:discussion}
\noindent
\textbf{System Design.}
A high-level idea of our system architecture is as follow:
initially, the AP knows the location of \shortname{} embedded on the wall. 
If the signal strength of the link between the AP and user is weak, 
the AP redirects its beam to \shortname{}. 
Then, \shortname{} generates multi-armed beams and concurrently steer them to scan the space
and finds the best path to the user (refer to~\cite{hassanieh2018fast} for beam-alignment using multi-beams). 
\shortname{} then reconfigures itself to form one beam with a maximum gain and 
steers the beam towards the user for data communication. 
In multi-user scenario, \shortname{} divides the beam and steer one beam to each user for multi-cast.
Bluetooth, $2.4$ GHz, or sub-$6$ GHz band can be used to exchange control information between the AP and \shortname{} and
to initially discover the AP at the end user side if the user is not in the same room as the AP.
When there are multiple rooms adjacent to a room where the AP is located, 
we may deploy a single \shortname{} on each wall between the adjacent rooms. 

\noindent
\textbf{Uplink Relay.}
We have previously discussed having metasurface that relays the AP signal to the user on the downlink. 
However, a single \shortname{} can act as either a downlink relay or a uplink relay.
Said differently, a wave propagating in the reverse direction interacts with 
the electric and magnetic response induced by \shortname{} in the same way as a wave propagating in the downlink direction. 
Thus, by properly applying the phase shifts to the uplink signal, we can steer the incident beam towards the AP.  

\noindent
\textbf{Implementation Cost.}
Given that mmWave links require LoS communication, 
we can assume that every room needs at least two mmWave APs 
for a full-coverage.
With two rooms, the number of the AP scales to four, and with three rooms, the number of the AP scales to six.
Since the state-of-the-art Wi-Fi 6 routers cost roughly $\$500$~\cite{netgear},
the implementation cost reaches approximately $\$2k$ for two rooms and $\$3k$ for three rooms and so on.
Using a smart metasurface, two rooms require only one \shortname{} and 
three rooms need two \shortname{} on the wall between the adjacent rooms.
We can breakdown the cost-driving factors of \shortname{} into the Roger4003C substrate, which costs approximately $\$300$~\cite{spwindustrial}, and varactor, which costs $\$2$ each~\cite{macom},
Therefore, we can assume that the implementation cost of \shortname{} is cheaper than deploying multiple mmWave APs.

\noindent
\textbf{Power Consumption.}
Metasurface consumes much less power than conventional array system that leverages patch arrays and phase shifters.
It is because metasurface eliminates 
the need for complicated RF chains, power amplifier, baseband processor, and/or active phase shifters. 
Said differently, the radiation pattern 
emitted by \shortname{} is the superposition of the field transmitted from
many meta-atoms, whose amplitude are
determined by the tunable resonance response of
the meta-atoms and the phase accumulated by their transmitted fields.
Since \shortname{} tunes each meta-atom using simple electric components like
varactors~\cite{macom}, it requires minimal additional power for beam
steering~\cite{POOLE2016519}.  In contrast, many conventional array antennas 
employ active phase shifters, which consume significant power~\cite{shlezinger2020dynamic}.

\section{Conclusion}
This paper presents \shortname{}, a smart surface system that reconfigures itself to relay the beam as a "lens" or "mirror" at the walls, provide abundant path options to remedy mmWave blockage, speed-up the beam alignment process, and to support multiple users, simultaneously. We also present promise for its capability through 
comprehensive experiments in HFSS simulation.
The results demonstrate that \shortname{} metasurface steers the outgoing 
beam in a
full $360$-degrees,  with an $89.8\%$ single-beam efficiency 
and $74.5\%$ double-beam efficiency.
We will fabricate this design 
to build an real smart surface prototype,
and focus on refining its system architecture. 

\section{Acknowledgements}
This work is supported by the National Science Foundation under grant CNS-1617161.

\bibliographystyle{abbrv} 
\begin{small}
\bibliography{hotmobile21}
\end{small}
\clearpage


\end{document}